\documentclass[aps,preprint,12pt,superscriptaddress,unsortedaddress,floatfix,bibnotes,nofootinbib,unsortedaddress]{revtex4}

\usepackage{epsfig}
\usepackage{subcaption}
\usepackage{color}
\def\vr{\mbox{\bf r}}

\def\kJm{~kJ~mol$^{-1}$}

\newcommand{\eqref}[1]{(\ref{#1})}

\newcommand{\kjm}{{kJ$\,$mol$^{-1}$}}

\begin{document}

\title{Adsorption of molecular hydrogen on honeycomb ZnO monolayers: A quantum density-functional theory perspective}

\author{ A.~Mart\'{\i}nez-Mesa}
\author{ L.~Uranga-Pi\~na}
\affiliation{Laboratoire Collisions Agr\'egats R\'eactivit\'e (FeRMI), Universit\'e Toulouse III - Paul Sabatier, UMR 5589, F-31062 Toulouse Cedex 09, France}
\affiliation{DynAMoS (Dynamical processes in Atomic and Molecular Systems), Facultad de F\'isica, Universidad de la Habana, Cuba}
\author{ N.~Halberstadt}
\affiliation{Laboratoire Collisions Agr\'egats R\'eactivit\'e (FeRMI), Universit\'e Toulouse III - Paul Sabatier and CNRS, UMR 5589, F-31062 Toulouse Cedex 09, France}
\author{ S. N.~Yurchenko}
\affiliation{Department of Physics and Astronomy, University College London, Gower Street, London WC1E 6BT, United Kingdom}
\author{T.~Heine}
\author{G.~Seifert}
\affiliation{Technische Universit\"at Dresden, Institut f\"ur Physikalische Chemie, D-01062 Dresden, Germany}

\begin{abstract}
We investigate the adsorption of molecular hydrogen on pristine zinc oxide (ZnO) platelets. The volumetric and gravimetric hydrogen storage capacities of the ZnO monolayers are evaluated in a broad range of thermodynamic conditions (i.e., for temperatures in the range 77~K $\le T \le$ 450~K, and for external gas pressures up to 200 bar). The thermodynamic properties and the microscopic spatial distribution of the adsorbed hydrogen fluid are assessed within the density functional theory of liquids for quantum fluids at finite temperature (QLDFT), and the adsorption enthalphies are obtained by fitting the computed adsorption densities to the Toth model isotherm. Compared to graphene platelets, the ZnO sheets impose a rather tighter confinement to the motion of the hydrogen molecules parallel to the surface. The isosteric heat of adsorption approaches 3.2~\kJm{} in the low density regime. This quantity shows a fairly smooth dependence on the hydrogen uptake for temperatures below 100~K, while it is shown to depend quite sensitively on the adsorbate density above this temperature.
\end{abstract}

\maketitle

\section{Introduction}
Understanding the fundamental mechanisms underlying nanoscale phenomena is one of the primary goals of modern physics and chemistry. This microscopic knowledge may assist in the design of advanced materials, tailored for specific applications, for instance in the fields of renewable energy production and storage, and of molecular electronics. In particular, there is a keen interest in the study of the adsorption properties of nanomaterials, owing chiefly to the potential applications in the design and fabrication of more efficient nanoscale devices for, e.g., hydrogen storage, nocive gas capture, sieving and sensing \cite{baig21}.

Molecular hydrogen is a renewable and environmentally friendly energy carrier. It is envisaged to play a key role in the energy transition, and to meet the ongoing increase of global energy demand. However, its large scale exploitation is hindered at present by the lack of efficient, lightweight hydrogen storage technologies, and the high costs of hydrogen production from renewable sources \cite{shen21}. In this context, hydrogen adsorption in lightweight nanostructured materials has been identified as a suitable alternative to meet current U.S. Department of Energy (DoE) hydrogen storage targets for onboard applications \cite{doe2022}.

Owing to its remarkable properties (e.g., thermal stability, low molecular weight, enhanced heat and electrical conductivities), graphene became the archetype of nanostructured materials shortly after its discovery in 2004 \cite{novoselov04}. It stimulated the synthesis of a wide variety of nanomaterials with well-defined structures, amenable for controlled fabrication and functionalization, and it triggered the current race for the development of two-dimensional materials for nanotechnology applications \cite{novoselov04,geim07,zendg18}.

As in many other energy-related areas (such as rechargeable batteries, supercapacitors for energy storage, solar cells), graphene-based nanomaterials have been intensively investigated regarding its ability to take up hydrogen. Although they exhibit lower hydrogen storage capacities, compared to more complex nanostructures such as metal-organic and covalent-organic frameworks, graphene-like nanomaterials remain as widespread models for the study of the adsorption in  more tangled nanomaterials.

One- to a few-atom thick ZnO sheets, similar to graphene, were theorised first \cite{topsakal09,si11,tu10,claeyssens05,freeman06,behera12} and successfully prepared afterwards \cite{young15,dahiya15,nazir22,sahoo16}. ZnO monolayers, like other zinc-oxide nanostructures (e.g., wires, belts, rings), attract a strong interest due to its potential applications in photocatalysis, and in the fabrication of light emitting diodes, field-effect transistors, supercapacitors, gas sensors, among others \cite{zhang22,young15,wakhare19,li17,borysiewicz19,hu09,balmeo21,zhang18,chen19,sun17,qu20,zhang17,tan16,qin14,dutta09}.

Moreover, ZnO monolayers may be regarded as a structural intermediate between paradigmatic examples of carbon nanomaterials like graphene, and of isoreticular metal-organic frameworks (MOFs) such as MOF-5. This analogy enables to use graphene-like ZnO sheets as a model system to provide insight into the fundamentals of hydrogen adsorption in more complex materials such as metal-organic frameworks. For example, it allows to assess the influence of structural parameters such as the specific surface area, and of atomic-level heterogeneities of the adsorbing surface, on the hydrogen storage capacity, by isolating them from other factors that also influence the hydrogen uptake by MOFs such as the porosity and the curvature of the inner surface \cite{hentsche07,han05,kadono03}.

The computational modelling and optimization of the physico-chemical properties of nanoscale matter is a key tool to tailor advanced materials for specific applications such as efficient and lightweight nanodevices for hydrogen storage. Specifically, sizeable quantum effects on hydrogen adsorption isotherms have been reported by previous computational investigations of H$_2$ physisorption on a variety of nanomaterials \cite{JChemPhys135,Thomas14,Kowalczyk15,Kagita12,Niimura12}.

The de Broglie thermal wavelength $\Lambda = \sqrt{ 2\pi \beta \hbar^2 /m }$ allows a simple assessment of the quantum delocalization of a system of structureless particles of mass $m$ in thermodynamic equilibrium ($\beta = 1/k_B T$ is the inverse temperature, and $k_B$ is the Boltzmann constant). The thermophysical properties of fluids at density $\rho$ can be accurately described within the framework of classical statistical mechanics, provided $\Lambda$ is much smaller than the average nearest neighbour separation $\rho^{-1/3}$ (that is, $\rho^{1/3}~\Lambda \ll 1$). This condition holds, for example, for a uniform hydrogen fluid at liquid nitrogen temperature ($T$=77 K) and at room temperature ($T$=298 K), as shown in Table \ref{table1}. However, adsorption in nanoporous materials results in an inhomogeneous distribution of hydrogen molecules, and local densities may increase up to values comparable to the density of liquid hydrogen ($\rho_{liq}$=70.8 kg/m$^3$ \cite{Leachman09}). At these concentrations, quantum effects are no longer negligible in the range of temperatures of interest for hydrogen storage (see Table \ref{table1}).

\begin{table}[htp!]
  \centering
  \caption{Ratio of the de Broglie thermal wavelength to the nearest neighbour separation, $\rho^{1/3}~\Lambda $, at $T$=77 K and at $T$=298 K. Densities $\rho (p=1~{\rm bar},T)$ and $\rho_{liq}$ are evaluated using the empirical equation of state for the normal hydrogen fluid \cite{Leachman09}.}
  \label{table1}
  \begin{tabular}{|c|c|c|} 
     \hline
       & $T$=77 K & $T$=298 K \\
     \hline
     $\rho (p=1~{\rm bar},T)$ & 0.06 & 0.02 \\
     \hline
     $\rho_{liq}$ & 0.37 & 0.19\\
     \hline
  \end{tabular}
\end{table}

Moreover, quantum  delocalization of the adsorbed hydrogen molecules is governed by the shape of the confinement potential imposed by the host structure (at low hydrogen uptakes), and by the superposition of the host-guest interaction potential and the intermolecular interactions (at high densities). Therefore, a reliable description of hydrogen storage in nanostructures over wide ranges of temperatures and pressures requires to account for quantum many-body effects on adsorption.

On the one hand, although the multidimensional Schr\"odinger equation enables the rigorous evaluation of both stationary and time-dependent properties of nanoscale systems, its numerical solution becomes computationally very demanding for systems with many degrees of freedom \cite{meyer09,marquardt20}. On the other hand, liquid density-functional theory provides an alternative, variational formulation of statistical mechanics \cite{kalikmanov01}, which enables the computer simulation of thermophysical properties of atomic and molecular fluids in thermodynamic equilibrium, at a fraction of the computational cost of standard numerical implementations of the path integral formulation of quantum mechanics \cite{shiga18}.  

The purpose of this paper is to present numerically converged calculations of molecular hydrogen adsorption on model ZnO monolayers, explicitly incorporating quantum confinement effects and intermolecular interactions within the numerical implementation of the Quantum Liquid Density Functional Theory (QLDFT) for molecular hydrogen at finite temperatures \cite{PhysRevE80}. 

The paper is organized as follows. The theoretical methodology employed in the QLDFT calculations of the  hydrogen uptake and the spatial distribution of hydrogen molecules on ZnO sheets, together with the relevant intermolecular interactions are presented in Section~\ref{sec:method}. Numerical details of the simulations are given in Section~\ref{sec:comp}. In Section~\ref{sec:results}, the results concerning the thermodynamic properties of the adsorbed fluid, the hydrogen storage capacities, and the microscopic density profiles, are presented and discussed. Finally, the main conclusions are laid out in Section~\ref{sec:summ}.

\section{Methodology}
\label{sec:method}

\subsection{Quantum Liquid Density Functional Theory}\label{ssec:qldft}

In the liquid density-functional formalism, as applied to atomic and molecular fluids, the thermodynamic potential is written as a functional of the density,
\begin{equation}
\Omega[\rho]=F[\rho]-\mu \int \rho(\vr)d^3 \vr. \label{eq:Omega} 
\end{equation}

% !!! Definir F

QLDFT is specifically tailored for the description of inhomogeneous hydrogen fluids \cite{PhysRevE80}. Within this framework, the effective Hamiltonian operator of the reference fluid reads 
\begin{widetext}
\begin{equation}
\label{e:H_s}
\hat H_s=-\frac{\hbar^2}{2M_{H_2}}\nabla^2+v_{ext}(\vr)+\int v_{12}(\vert
\vr-\vr' \vert)\rho(\vr')d^3 \vr'+v_{xc}[\rho
(\vr) ].
\end{equation}
\end{widetext}

The reference fluid is composed by non-interacting particles with identical mass $M_{H_2}$ to that of hydrogen molecules, and moving on the effective potential $v_{eff}\left[ \rho(\vr) \right]$ which involves three different contributions, namely the external potential $v_{ext}$ exerted on hydrogen molecules by the host nanostructure, the mean-field potential due to the H$_2$-H$_2$ interaction $v_{12}$, and the so-called exchange-correlation contribution $v_{xc}=\delta F_{xc}/\delta \rho(\vr)$. Particles in the reference fluid build up the same density $\rho(\vr)$ as the real, interacting H$_2$ molecules. Hereafter, the density $\rho(\vr)$ is normalized to the average number of particles $N$.

The Helmholtz free energy of the system is given by
\begin{equation}
F[\rho]=F_s[\rho]-\frac{1}{2} \int \int \rho(\vr)\rho(\vr')v(\vr-\vr')d^3 \vr d^3 \vr'-\int
\rho(\vr)v_{xc}(\vr)d^3 \vr+F_{xc}[\rho], \label{eq:FpT} 
\end{equation}
where $F_s[\rho]$ is the free energy of the reference fluid. The second and third terms are corrections introduced to avoid double counting of the interaction energy due to mean-field and exchange-correlation contributions, respectively.

In the density-functional calculations, the internal structure of the H$_2$ molecules is disregarded (a detailed assessment of the consequences of this approximation can be found elsewhere \cite{PNAS102,PCCP9}). Thus, the intermolecular potential $v(\vert \vr-\vr' \vert)$ depends only on the relative distance between the molecular centres of mass. The exchange-correlation contribution to the free energy, $F_{xc}[\rho]$, is derived from the empirical data available for the homogeneous fluid, $F_{expt}[\rho,T]$ \cite{JChemPhys66,Leachman09}, within the Local Density Approximation, as follows:
\begin{equation}
F_{xc}[\rho]=\int \rho(\vr)\epsilon_{xc}(\vr)d^3 \vr ~~~ \Rightarrow ~~~ v_{xc}(\vr)=\epsilon_{xc}[\rho(\vr)]+\rho(\vr)\frac{\delta
\epsilon_{xc}}{\delta \rho(\vr)} \label{eq:vxc},
\end{equation}
whereas the exchange-correlation energy density given by
\begin{equation}
\epsilon_{xc}[\rho]=\frac{F_{expt}[\rho,T]}{N}+\frac{1}{\beta}\log \left(
\frac{Z_{kin}}{\rho V} \right)-\frac{\rho}{2} \int v_{12}(\vert \vr-\vr'
\vert)d^3 \vr' 
\label{eq:exc}. 
\end{equation}
$Z_{kin}$ is the single-particle canonical partition function in the free space. The empirical equation of state holds for pressures up to 2~GPa, which is sufficient for the investigation of hydrogen adsorption on ZnO monolayers.

The spatial distribution $\rho(\vr)$ of the hydrogen molecules is given by the diagonal elements of the number operator $\hat \rho_s$. The latter is evaluated as a series expansion of the Bose-Einstein distribution function \cite{JChemPhys135,JPhysChemC116}, using a three-dimensional finite-difference representation of the effective Hamiltonian $\hat H_s$.
\begin{equation}
\hat \rho_s = \frac{1}{e^{ \beta \left( \hat H_s-\mu \right)}-1} =
\sum_{k=1}^{\infty} e^{ -k\beta \left( \hat H_s-\mu \right)} =
\sum_{k=1}^{\infty} \sum_{m=0}^{\infty} \frac{\left[ -k\beta \left( \hat
H_s-\mu \right) \right]^m}{m!} ,
\label{eq:pow_rho}
\end{equation}
where $\beta=1/k_BT$ is the inverse temperature. 

The average concentration $n_{\rm{ads}}$ of adsorbed molecules (confined to a volume $V$ and in equilibrium with an external hydrogen gas at the temperature $T$ and pressure $p$) is computed as the trace of the number operator $\hat \rho_s$,
\begin{equation}
n_{\rm{ads}} = \frac{1}{V} \int \rho(\vr) d^3 \vr, \label{eq:nads}
\end{equation} 
and subsequently used to assess the equilibrium constant $K_e$, and the gravimetric storage $gw$ capacity of the nanostructured surface:
\begin{eqnarray}
K_e & = & n_{\rm{ads}}/n, \label{eq:Ke} \\
gw (\%) & = & \frac{100\cdot n_{\rm{ads}}}{n_{\rm{ads}}+\frac{M_{Zn}}{M_{H_2}}n_{Zn}+\frac{M_O}{M_{H_2}}n_O}, 
\label{eq:gw}
\end{eqnarray} 
where $n$ is the density of the external gas, while $M_{Zn}$ ($M_O$) and $n_{Zn}$ ($n_O$) represent the the mass of a zinc (oxygen) atom and the number density of zinc (oxygen) atoms in the host structure, respectively.

In equation (\ref{eq:gw}), $gw$ stands for the {\it total} gravimetric hydrogen storage capacity of the host surface. If only the difference between the density of adsorbed molecules $n_{\rm{ads}}$ and the bulk H$_2$ density $n$ at the same temperature and pressure is considered (i.e., $n_{\rm{ads}} \Rightarrow n_{\rm{ads}} - n$ in the numerator of equation (\ref{eq:gw})), the {\it excess} gravimetric storage capacity is computed instead.

The grand potential $\Omega$ of the interacting system was evaluated from equations (\ref{eq:Omega}) and (\ref{eq:FpT}), i.e., by subtracting double counting corrections from the grand potential of the reference fluid:
\begin{eqnarray}
 \Omega_s & = & {\rm Tr} \left\lbrace \hat \Omega_s \right\rbrace ,\\
\hat \Omega_s & = & -\frac{1}{\beta}\log Z_s = -\frac{1}{\beta} \sum_{k=1}^{\infty}
\sum_{m=0}^{\infty} k^{m-1} \frac{ \left[ -\beta \left( \hat H_s-\mu \right)
\right]^m }{m!} .
\label{eq:pow_logZ}
\end{eqnarray}

\subsection{Isosteric enthalpy of adsorption}\label{ssec:qst}

To charaterize the thermal performance of ZnO sheets as hydrogen adsorptive media, the isosteric adsorption enthalpy $Q_{st}$ was evaluated from the H$_2$ uptake isotherms (computed at selected temperatures in the range between 77~K and 450~K). To this purpose, we employ the Clausius-Clapeyron relation:
\begin{equation}
 Q_{st} = RT^2\left( \frac{\partial \log p}{\partial T} \right)_{n_{ads}}, \label{eq:qst}
\end{equation}
were $R$ is the universal gas constant, and derivatives are taken at constant adsorbate concentration. $Q_{st}$ values computed using equation (\ref{eq:qst}) are very sensitive to small variations of the input data, or to the functional form assumed in isotherm fit models \cite{Siperstein21}. Therefore, instead of evaluating numerical derivatives directly from the calculated volumetric capacities, we have chosen to fit the data using a Toth model isotherm \cite{Toth71}:
\begin{equation}
 n_{ads} = \frac{n_{sat}bp}{\left[ 1+\left( bp \right)^t \right]^{1/t}}. \label{eq:toth}
\end{equation}
The Toth isotherm is often used to model adsorption on heterogeneous substrates, i.e., in materials with various types of adsorption sites and  distinct binding energies \cite{Vasanth11,Purewal12,Kloutse15,Chilev22}. It overcomes limitations of other models routinely used to describe adsorption on heterogeneous surfaces, such as the Freundlich \cite{Freundlich06} or Sips \cite{Sips48} isotherms, since both the low pressure limit of the isotherm slope and the saturation uptake are well defined for the Toth functional form. In equation (\ref{eq:toth}), $n_{sat} = n_{ads}(p\rightarrow \infty, T)$ represents the hydrogen uptake at saturation, $b$ is the Henry adsorption constant or affinity, and $t$ is the heterogeneity parameter. 

Inserting the function (\ref{eq:toth}) in equation (\ref{eq:qst}), and assuming that the heterogeneity parameter depends linearly on temperature (that is, $t = t_0 + \kappa T$, which fits the data quite well (see section \ref{sec:results})), the expression for the isosteric heat can be cast in the following form:
\begin{equation}
 Q_{st} = -RT^2 \left\lbrace \frac{d \log b}{dT} + \left[ 1+\left( bp \right)^t \right] \frac{d \log n_{sat}}{dT} - \frac{\kappa}{t} \left[ \log \left( \frac{n_{ads}}{n_{sat}} \right) \left[ 1+\left( bp \right)^t \right] - \log \left( bp \right) \right] \right\rbrace .
\end{equation}

\section{Computational details}
\label{sec:comp}

\begin{figure}[htp!]
\begin{center}
\includegraphics[width=0.65\textwidth]{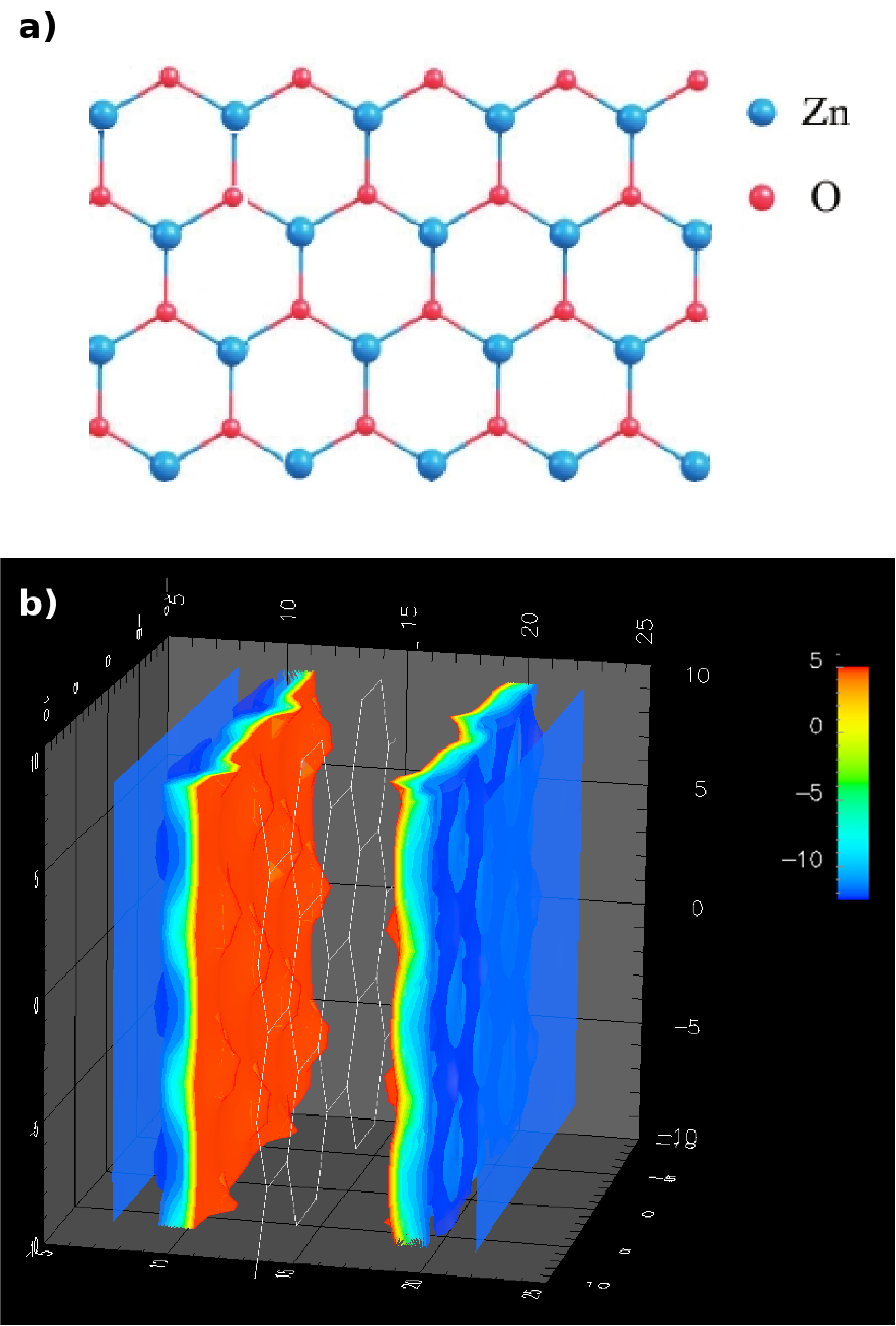}
\caption{\label{fig:vext}(a) Atomic positions in the ZnO monolayer. (b) H$_2$-surface interaction potential (energies in \kJm, distances in atomic units).}
\end{center}
\end{figure}

ZnO monolayers constitute lightweight nanomaterials with large specific surface area (1390~m$^2$/g). Figure \ref{fig:vext} displays the geometrical arrangement of atoms in a ZnO sheet, as well as the interaction potential $v_{ext}$ acted on guest H$_2$ molecules by the substrate.

The description of the interaction potential between the H$_2$ molecules and the host surface used in the present simulations is based on the Optimized Potentials for Liquid Simulations (OPLS) force field developed by Jorgensen et al. \cite{Jorgensen96}. The parameters of the non-bonded contributions in the OPLS model were obtained using geometric combining rules (i.e., $\varepsilon_{\alpha \alpha'}=\sqrt{\varepsilon_{\alpha \alpha}~\varepsilon_{\alpha' \alpha'}}$, and $\sigma_{\alpha \alpha'}=\sqrt{\sigma_{\alpha \alpha}~\sigma_{\alpha' \alpha'}}$, with $\alpha, \alpha'= \left\lbrace H,~Zn,~O \right\rbrace $), and the homonuclear atomic van der Waals energies $\varepsilon_{aa}$ and distances $\sigma_{aa}$ were taken from the literature \cite{Yang05}. 

The choice of the OPLS force field is based on the assumption that the Lennard-Jones functions provide a reasonable description of the decay of dispersion contributions with the distance, and the parameterization introduced in Ref. \cite{Yang05} is readily extensible to other nanomaterials such as metal-organic frameworks. Moreover, the estimated adsorption properties of the material are expected to be more sensitive to the overall well depth of the interaction potential between the molecules and the adsorptive surface. When considering hydrogen molecules as point particles, the binding energy predicted by the present OPLS model (13.5\kJm) is in good agreement  with the results of previous DFT calculations of H$_2$ adsorption on honeycomb ZnO sheets (13.2\kJm) \cite{si11}.

Contour plots of the H$_2$-host potential energy surface, depicted in figure \ref{fig:vext}, were obtained by taking into account the contributions to the interaction potential of three additional unit cells along each direction. The inclusion of further replicas of the unit cell does not appreciably modify $v_{ext}(\vr)$.  It can be seen that the main adsorption sites are separated by about 5 $a_0$ from the substrate, and they are preferentially located on top of the Zn and O atoms, and of Zn-O bonds. The hydrogen binding energy (increased roughly by 50\% with respect to the H$_2$-graphene interaction) decreases only slightly between neighbouring main adsorption sites, therefore the latter reproduce the hexagonal pattern of the honeycomb lattice.

The intermolecular hydrogen-hydrogen interactions were modelled through a Morse function 
\begin{equation}
 v_{12}(r) = D \left[ e^{-2\alpha (r-r_e)} - 2e^{-\alpha (r-r_e)} \right]
\end{equation}
with parameters $D$=0.291\kJm, $r_e$=3.511~\AA{} and $\alpha$=1.592~\AA{}$^{-1}$. The potential parameters were obtained from the fit of the isotropic average of {\it ab initio} calculations of the intermolecular potential energy surface \cite{JChemPhys113}.

In QLDFT calculations, upon inclusion of the surface-hydrogen, hydrogen-hydrogen and exchange-correlation contributions to the effective interaction potential $v_{eff}$, a cut-off of 42\kJm{} above the global minimum was introduced in order to improve the numerical stability of the algorithm (for details see \cite{PhysRevE80}). QLDFT simulations of hydrogen physisorption on ZnO monolayers were systematically conducted at the LIE-1 level of theory \cite{PhysRevE80}, and the results are reported in section \ref{sec:results}.

We considered a broad range of thermodynamic conditions relevant for automotive applications, i.e., temperatures from the nitrogen condensation point ($T_{min}$=77~K) to $T_{max}$=450~K, and a range of external gas densities between a dilute gas ($n_{min}$=5.4$\cdot$10$^{-4}$~kg/m$^3$, corresponding to a molar volume $v_m$=3.74~m$^3$) and $n_{max}$=55.6~kg/m$^3$ ($v_m$=36.2~cm$^3$). As a reference, this density range spans from 0.6\% to more than 600\% of the hydrogen density at normal temperature and pressure $n_{n.t.p.}$=8.4$\cdot$10$^{-2}$~kg/m$^3$ ($v_m$=24.1~dm$^3$) \cite{Leachman09}.

As the exchange-correlation potential and its contribution to the free energy, the chemical potential $\mu$ is retrieved, for every external gas pressure $p$ and temperature $T$, from the experimental data of the homogeneous fluid:
\begin{equation}
 \mu = \frac{F_{expt}(p,T)}{N}+\frac{p}{n} \label{eq:mu}
\end{equation}
As stated above, the matrix elements of the occupation number operator $\hat \rho_s$ are calculated via the formal power expansion (\ref{eq:pow_rho}) by introducing a finite-difference representation of the Hamiltonian in a three-dimensional grid. The number of points in the uniform mesh was varied in order to keep the coarsest grid spacing along each of the Cartesian axis below 0.4~\AA{}, which is found to yield converged results. The QLDFT set of equations (\ref{eq:pow_rho}), (\ref{eq:vxc}), (\ref{eq:exc}) and (\ref{e:H_s}), are solved self-consistently by damped stationary point iterations 
\begin{eqnarray}
\rho^{(i)}(\vr)=\lambda \rho^{(i)}(\vr)+(1-\lambda)\rho^{(i-1)}(\vr)\\
v_{eff}^{(i)}(\vr)=\lambda v_{eff}^{(i)}(\vr)+(1-\lambda)v_{eff}^{(i-1)}(\vr)
\end{eqnarray}
with a mixing coefficient of $\lambda=0.15$. Iterations start from the density field $\rho_{clas}(\vr)$ predicted by the classical liquid density functional theory for the same external potential \cite{JChemPhys135,JPhysChemC116}. The series expansions (\ref{eq:pow_rho}) and (\ref{eq:pow_logZ}) were cut off at $k_{max}=5$ and $m_{max}=50$, since those terms were sufficient to achieve convergence in the whole range of temperatures and pressures considered. At every step, matrix elements smaller than $10^{-4}$ times the largest value were neglected in the computation of $\hat \rho_s$.

\section{Results and discussion}
\label{sec:results}

In the following, we present the results of the QLDFT simulations for the thermodynamic and structural properties of molecular hydrogen physisorbed on ZnO sheets, namely the average binding free energy of H$_2$ molecules, the density of adsorbed molecules (volumetric storage capacity), the gravimetric storage capacity, and the density distribution of the hydrogen fluid.

\subsection{Thermophysical properties of the adsorbed hydrogen fluid}

\begin{figure}[htp!]
\begin{center}
\includegraphics[width=0.95\textwidth]{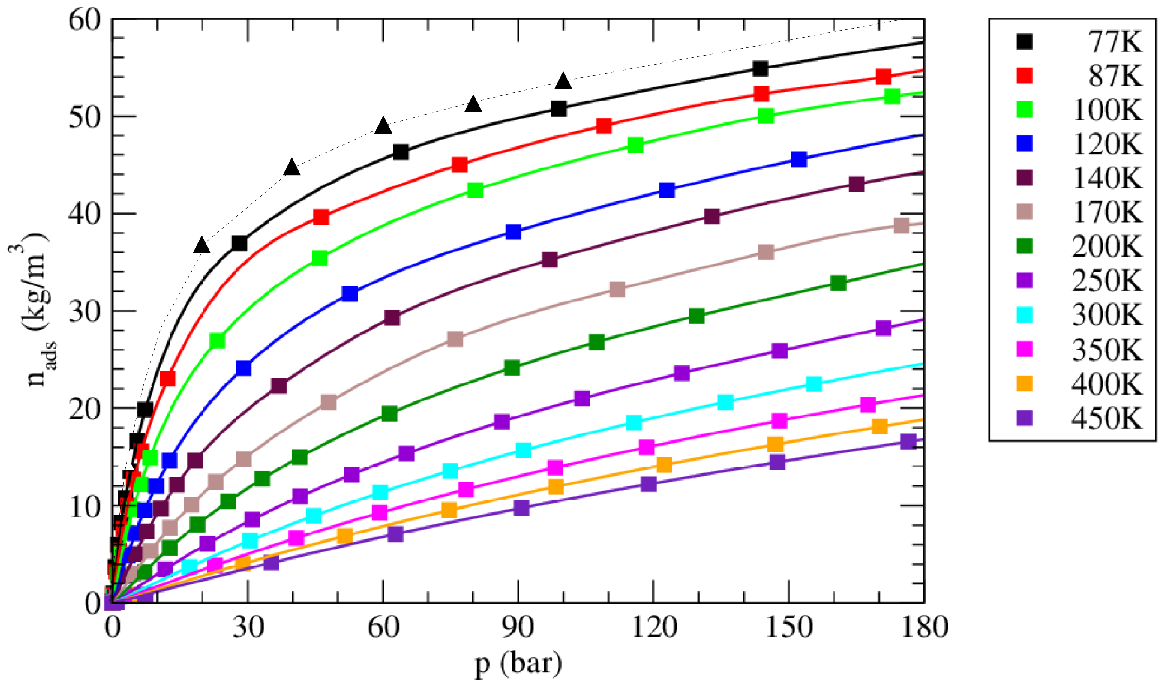}
\caption{\label{fig:v-capacity}Volumetric storage capacity: Computed adsorption isotherms n$_{\rm{ads}}(p,T)$ for H$_2$ on ZnO sheets, at selected temperatures between 77 K and 450 K, and for external gas pressures up to 180~bar. The up triangles represent the results of previous QLDFT calculations for the density of adsorbed molecules in MOF-5 at $T$=77~K \cite{Thomas14}.}
\end{center}
\end{figure}

Hydrogen adsorption isotherms on zinc oxide monolayers are shown in Figure~\ref{fig:v-capacity}. It can be seen, that the computed $n_{\rm{ads}}(p,T)$ depends sensitively on the thermodynamic conditions of the external gas. A single ZnO sheet can take up to approximately 4~kg/m$^3$ of hydrogen within the range of pressures and temperatures defined by the US DoE targets for operation of hydrogen storage devices in automotive applications, i.e., at pressures below the ``maximum delivery pressure'' ($p^{(DOE)}_{~target}$=12~bar) and above the ``operating ambient temperature'' and ``minimum delivery temperature'' ($T^{(DOE)}_{target}$=233~K) \cite{doe2022}. Meanwhile, notably larger adsorbate concentrations can be achieved if more flexible operating conditions are acceptable, e.g., the DoE goals for the volumetric capacity of onboard hydrogen storage capacities are met on ZnO layers at moderate external gas pressures ($p\le$100~bar), respectively, for temperatures below 170~K (present target, 30~kg/m$^3$) and at liquid nitrogen temperature (ultimate target, 50~kg/m$^3$). 

Remarkably, the hydrogen adsorption isotherms at liquid nitrogen temperature, $n_{\rm{ads}}(p,T{\rm{=77~K}})$, look very much alike for zinc oxide monolayers and for MOF-5 \cite{Thomas14}. Owing to the difference in specific masses, a similar correspondence does not hold for the total gravimetric storage capacities of these two nanomaterials. Beyond the specific case of ZnO platelets and MOF-5, the equivalence between the hydrogen uptake in simple two-dimensional structures and in nanomaterials with more complex geometry or chemical composition (such as metal- and covalent-organic frameworks) is particularly appealing from the computational point of view. Indeed, more accurate electronic structure calculations are affordable for the former systems, which may be used as model systems to gain further insight into the microscopic properties underlying hydrogen storage capacities of the more intricate materials. 

\begin{figure}[htp!]
\begin{center}
\includegraphics[width=0.95\textwidth]{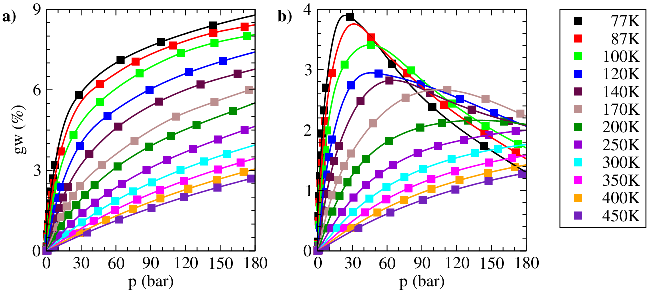}
\caption{\label{fig:g_capacity}a) Overall gravimetric capacity, and b) excess gravimetric capacity for H$_2$ on ZnO sheets, at selected temperatures between 77 K and 450 K, and for external gas pressures up to 180~bar.}
\end{center}
\end{figure}

Figure \ref{fig:g_capacity}a illustrates the pressure dependence of the gravimetric H$_2$ storage capacity of the ZnO layer. In the range of thermodynamic conditions defined by the US DoE targets (i.e., for $T\ge T^{(DOE)}_{~target}$ and $p\le p^{(DOE)}_{~target}$), the gravimetric capacity of zinc oxide surfaces is below 1\%. Hydrogen uptake exhibits steep increases at lower temperatures, in the low-pressure regime, attaining 4.5\% at liquid nitrogen temperature and at an external gas pressure $p$=12~bar (i.e., matching the current DoE target for the gravimetric capacity of storage media).

At $T$=300~K and $p$=50~bar, the adsorbed H$_2$ molecules amount to 1.6\% of the total weight of the system. The QLDFT prediction of the gravimetric capacity is close to the lower bound of previous theoretical estimations (1.5\%~$\le gw \le$~2.1\%)) based on an independent-particle picture of physisorption, for the same thermodynamic conditions \cite{si11}. This is consistent with the mild influence of the interaction between hydrogen molecules in neighbouring adsorption sites. On the one hand, at the corresponding adsorbate density ($n_{\rm{ads}}$=10~kg/m$^3$), storage properties at room temperature are primarily driven by the adsorption potential well depth. On the other hand, deviations from the linear behaviour (as a consequence of increasing surface coverage) are clear in the $gw(p,T{\rm{=300~K}})$ curve around $p$=50~bar. 

At moderately high pressures ($p\gtrsim$150~bar), the total gravimetric capacity spans between 3.5\% and 8.5\%, for temperatures between $T$=300~K and $T$=77~K.

In Figure \ref{fig:g_capacity}b, we also plot the excess gravimetric storage capacity of the target material, which is a quantity more amenable for direct experimental verification. Concomitantly, the excess storage capacity allows to characterize the influence of the nanostructure on the hydrogen uptake. For the range of thermodynamic conditions explored in this work, the highest excess storage capacity (3.9\%) is achieved for an external gas pressure of 25~bar, and at $T$=77~K.

At temperatures below 200~K, it is possible to define {\it optimal} operation points ($p_{opt}(T)$) based on the dependence of the excess gravimetric capacity on the external gas pressure (i.e., the thermodynamic conditions at which the adsorptive material causes a higher relative increase of the hydrogen uptake with respect to storage in an empty container). The maxima in the excess storage capacity (for instance, $p_{opt}\sim$25~bar at liquid nitrogen temperature, and $p_{opt}\sim$45~bar at $T$=100~K) arise from the competition between the opposite effects, on the molecular adsorption, of the intermediate region (that is, the potential well) and the short-range region of the H$_2$-surface interaction potential.

(i) In the low-coverage regime, it is energetically more favourable for H$_2$ molecules to occupy individual adsorption sites in the vicinity the ZnO surface, thus the presence of the structure provokes a local enhancement of hydrogen concentration compared to the density of the external gas;

(ii) In the high-pressure regime, beyond the limit of complete surface coverage, the main adsorption sites can not accommodate additional hydrogen molecules. The presence of the ZnO layer becomes counterproductive beyond the saturation pressure, since the overlap between the electronic densities of the adsorbate molecules and the atoms in the substrate leads to strongly repulsive forces at short range H$_2$-surface separations, thereby decreasing the effective accessible volume in comparison to an empty tank.

Conversely to total volumetric storage capacities plotted in figure \ref{fig:v-capacity}, the excess gravimetric capacity of ZnO monolayers at $T$=77~K differs from that previously reported for MOF-5. In ZnO sheets, the $gw(p,T{\rm{=77~K}})$ curve exhibits a sharp peak rather than the plateau-like behaviour predicted by QLDFT calculations of H$_2$ storage in MOF-5 \cite{Thomas14}. Moreover, for the ZnO layer, the change in the slope of the $gw(p,T{\rm{=77~K}})$ curve occurs at a lower pressure (25~bar) in contrast to the metal-organic framework (45~bar). These differences between the excess storage capacities of the two systems follow from the more complex H$_2$-substrate interaction potential in MOF-5, e.g., the presence of multiple adsorption sites, and the comparatively wider adsorption region. Incidentally, the pressure of 25~bar (at which the change in the slope of the computed $gw(p,T{\rm{=77~K}})$ curve is taking place for the zinc oxide nanosheet) matches very well the corresponding point in the pressure dependence of the excess gravimetric uptake in MOF-5, measured in a series of experiments \cite{kaye07,zhou07,wongfoy06,panella06,dailly06}.

The computed adsorption isotherms are used to evaluate the isosteric enthalpy $Q_{st}$ of hydrogen adsorption on ZnO sheets, as outlined in section \ref{sec:method}. The Toth model assumes an asymmetrical, quasi-Gaussian distribution of adsorption sites energies, and a preponderance of adsorption energies lower than the maximum of the distribution \cite{Vasanth11}. It fits very well the computed volumetric capacities $n_{ads}(p,T)$ (the lowest value obtained for the coefficient of determination, in the non-linear regression analysis, was 0.999858).

\begin{figure}[htp!]
\begin{center}
\includegraphics[width=0.9\textwidth]{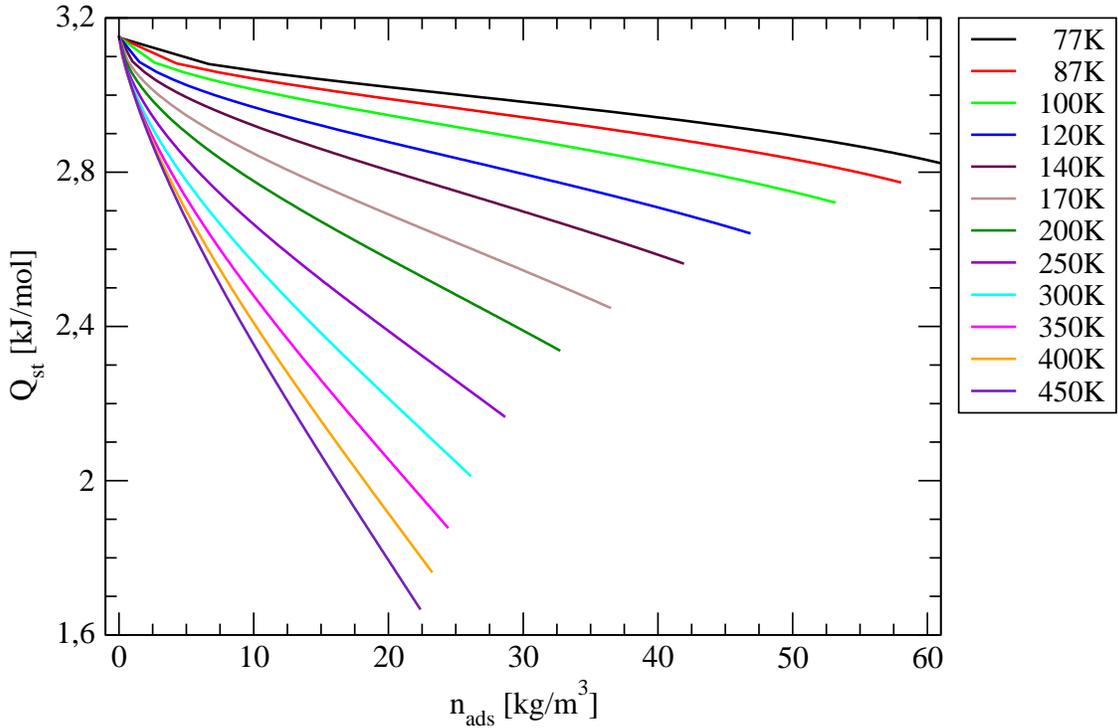}
\caption{\label{fig:binding_f}Isosteric heat of adsorption $Q_{st}$ of molecular hydrogen on a ZnO substrate, as a function of the density $n_{ads}$ of the adsorbed fluid, and for different temperatures between 77~K and 450~K.}
\end{center}
\end{figure}

The curves $Q_{st}(n_{ads},T)$ characterize the binding strength of guest hydrogen molecules to the layered material for varying thermodynamic conditions. They are plotted in figure \ref{fig:binding_f} as a function of the temperature and of the density of the adsorbed fluid. It can be seen, that while the isosteric heat of adsorption is nearly insensitive to changes in temperature for low surface coverage (i.e., $Q_{st}(n_{ads}\rightarrow 0,T) \approx$3.2~\kjm), this quantity exhibits a marked temperature dependence at higher adsorbate concentrations. The decay of the isosteric heat weighs the reduced propensity of the ZnO nanostructure to adsorb hydrogen for increasing temperatures.

Likewise, a monotonous decrease of $Q_{st}(n_{ads},T)$ occur for increasing hydrogen uptake. The observed decrease in the isosteric adsorption enthalpy as a function of adsorbate concentration (or, equivalently, of the external gas pressure) is milder for the lowest temperatures considered, and it becomes progressively steeper at larger temperatures. Indeed, the variations of $Q_{st}$ are lower than 30\% over the whole range of adsorbed hydrogen densities, for temperatures $T\le 100$~K, while the adsorption enthalphy decrease roughly linearly at room temperature and above. These results are in line with recent grand canonical Monte Carlo simulations and adsorption microcalorimetry experiments, which also predicted a significant temperature dependence of adsorption enthalpy for small molecules in microporous materials \cite{Rubes18,Hyla19,Siderius22}. Besides, since constant isosteric enthalpy of hydrogen adsorption is sought as a mean to simplify the engineering requirements for energy storage in onboard applications, these results support the conclusion that the potential use of ZnO monolayers to this purpose would be restricted to temperatures below liquid oxygen temperature.

\begin{figure}[htp!]
\begin{center}
\includegraphics[width=0.85\textwidth]{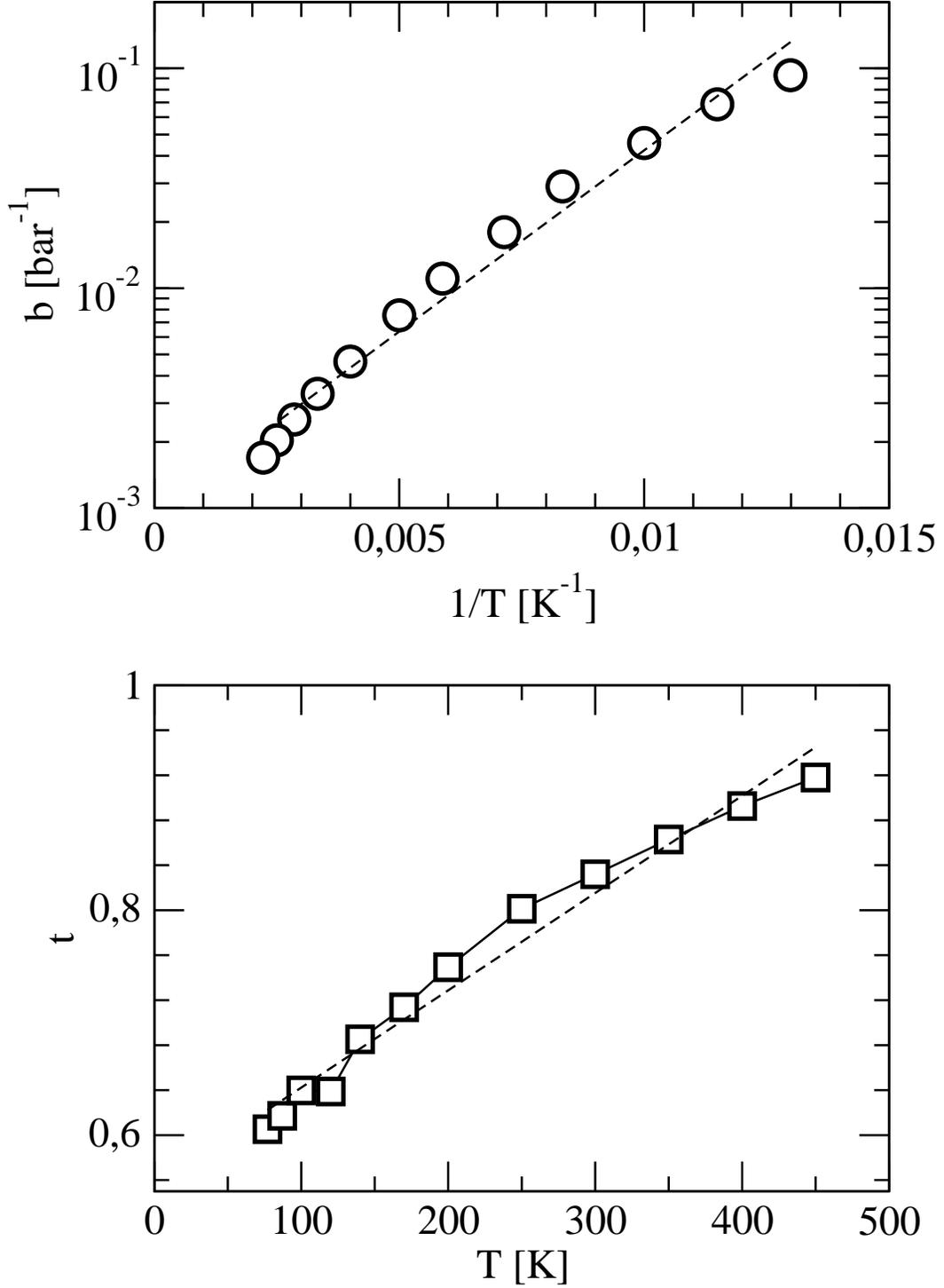}
\caption{\label{fig:henry}a) Henry constant $b$ for molecular hydrogen adsorption on ZnO sheets, as a function of inverse temperature. b) Heterogeneity parameter $t$, as a function of temperature. The straight (dashed) lines providing the best fits of the $b(T^{-1})$ and the $t(T)$ data (within the temperature ranges 77~K$\le T \le 250$~K and 77~K$\le T \le 450$~K, respectively) are included to guide the eye.}
\end{center}
\end{figure}

The van 't Hoff plot of Henry’s law constant is shown in figure \ref{fig:henry}a). The Henry constant characterizes the performance of the adsorptive media in the low pressure regime, and it is traditionally used as one of the key figures of merit in the computational screening of adsorbent materials \cite{Siderius22}. The computed $b$ values closely follows the van 't Hoff equation for temperatures up to 250~K, while in contrast deviations for linearity become noticeable above room temperature. The slope of the van 't Hoff plot provides an alternative route to determine the isosteric heat of adsorption, different from using the Clausius-Clapeyron and the uptake isotherms. Restricting the analysis to the linear part of the plot ($T\le 250$~K), we obtain $Q_{st}^{(vtH)} = 3.41$\kJm. This value is found to be in reasonable agreement with the results reported in figure \ref{fig:binding_f}, taking into account that the latter are inferred from the behaviour of adsorption isotherms over the whole range of temperatures and external gas pressures, and $Q_{st}^{(vtH)}$ is representative of adsorption in the low-coverage regime only.

The temperature-dependent heterogeneity factor $t$ is shown plotted in figure \ref{fig:henry}b). It confirms the assumed linear T-dependence of this parameter, and the heterogeneous character of the adsorption process (the coefficient $t(T)$ deviates from unity in most of the temperature range).

\subsection{Microscopic structure of the adsorbed hydrogen fluid}

\begin{figure}[htp!]
\begin{center}
\includegraphics[width=0.5\textwidth]{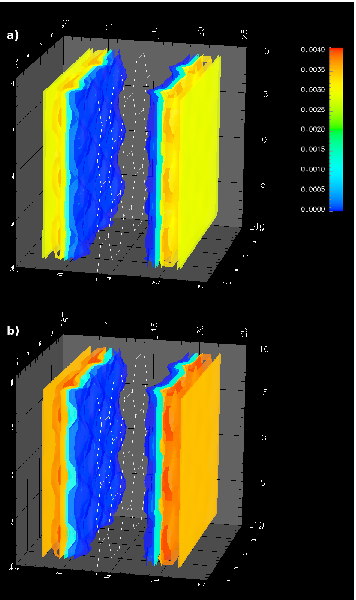}
\caption{\label{fig:density77K}Spatial distribution of hydrogen molecules in the vicinity of a ZnO monolayer, at the temperature $T$=77~K, and the pressure a) $p$=28~bar, and b) $p$=140~bar. The density is expressed in atomic units ($a_0^{-3}$).}
\end{center}
\end{figure}

The computed volumetric and gravimetric storage capacities suggest that ZnO monolayers can adsorb substantial quantities of hydrogen at cryogenic temperatures and under moderate pressure conditions. As an illustration of the density distribution of adsorbed hydrogen molecules under these thermodynamic conditions, the contour plots of the adsorbate density at the temperature $T$=77~K, and the external gas pressures $p$=28~bar and $p$=140~bar are depicted in figure \ref{fig:density77K}. 

The constant density surfaces show the formation of well localized density peaks in the direction perpendicular to the surface, whereas these peaks are poorly resolved in the directions parallel to the substrate. Nonetheless, the computed microscopic density profiles are more structured than the distribution of H$_2$ molecules adsorbed on graphene platelets, and other honeycomb carbon nanostructures \cite{JChemPhys135,PNAS102,JPhysChemC116,RCF2014,PCCP6,NanoLett1}. For graphene-like carbonaceous materials, the molecular density gets delocalized along the graphene planes, indicating free lateral motion of the molecules along the pore walls. Conversely, the present QLDFT simulations confirm that the local maxima in the spatial distribution of adsorbed hydrogen molecules are spread evenly on top of surface atoms, and of bridge sites (at both sides of the zinc oxide layer), thereby mimicking the hexagonal pattern of the substrate.

At $p$=28 bar, the density at the main adsorption sites amounts to 3.57$\cdot 10^{-3}a_0^{-3}$ (corresponding to a molar volume of 25 cm$^3$). These density peaks are 8.7 times higher than the average concentration of the homogeneous hydrogen fluid at the same temperature and pressure, and it translates in  the maximum excess gravimetric capacity reported in figure \ref{fig:g_capacity}.

While increasing the external gas pressure by a factor of five (up to 140 bar) leads to a 49\% increment of the overall hydrogen uptake, such gain is chiefly ascribable to enhanced adsorption beyond the main adsorption sites. As a matter of fact, the maximum local density at 140 bar (4$\cdot 10^{-3}a_0^{-3}$, corresponding to a molar volume of 22 cm$^3$) is only 12\% larger than the peaks of the adsorbate density at p=28 bar. Concurrently, the local maxima of the probability density at $T$=77~K and $p$=140 bar represent 2.2 times the density of homogeneous hydrogen at the same temperature and pressure. Hence, the distribution of the adsorbed hydrogen becomes less structured at larger pressures, and the excess gravimetric capacity of the nanostructure deteriorates.  

\begin{figure}[htp!]
\begin{center}
\includegraphics[width=0.5\textwidth]{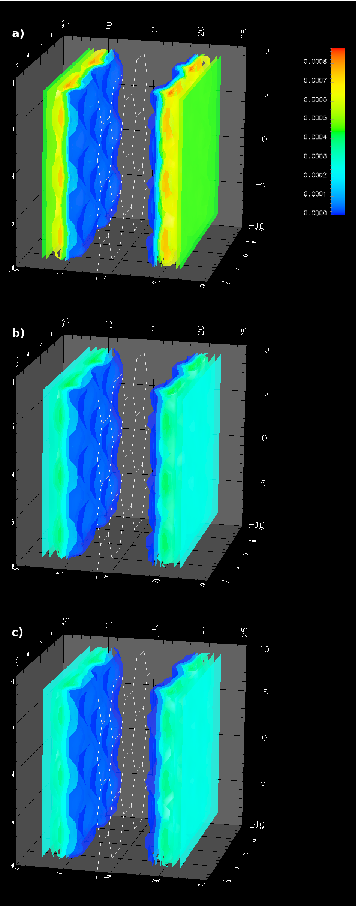}
\caption{\label{fig:density12bar}Density distribution of hydrogen molecules adsorbed on a ZnO surface, at pressure $p^{(DOE)}_{~target}$=12~bar and temperature a) $T$=200~K, b) $T$=250~K, and c) $T$=300~K. The density is expressed in atomic units ($a_0^{-3}$).}
\end{center}
\end{figure}

Figure \ref{fig:density12bar} displays the hydrogen density distribution in the vicinity of the solid surface at the pressure $p^{(DOE)}_{~target}$=12~bar, and for temperatures $T$=200~K, 250~K, and 300~K. Within this range of temperatures, hydrogen molecules also occupy primarily {\it top} and {\it bridge} sites on the ZnO lattice. At the main adsorption sites, the local maxima of the hydrogen density decreases from 8.8$\cdot 10^{-4}a_0^{-3}$ (equivalent to 19.9 kg/m$^3$) at 200 K down to 4.1$\cdot 10^{-4}a_0^{-3}$ (9.3 kg/m$^3$) at room temperature. It can be seen that raising the temperature from 200 K up to 250 K causes a steep reduction of the overall hydrogen uptake. A milder decline of the adsorbed density is observed as a consequence of further increase of the temperature up to 300 K. It hints at the suitability of ZnO based nanomaterials to store hydrogen at the DOE recommended ``operating ambient temperature'' (that is, T$\ge$200 K), provided the key figures of merit can be further improved, e.g., by chemical doping or geometry optimization.

\begin{figure}[htp!]
\begin{center}
\includegraphics[width=0.95\textwidth]{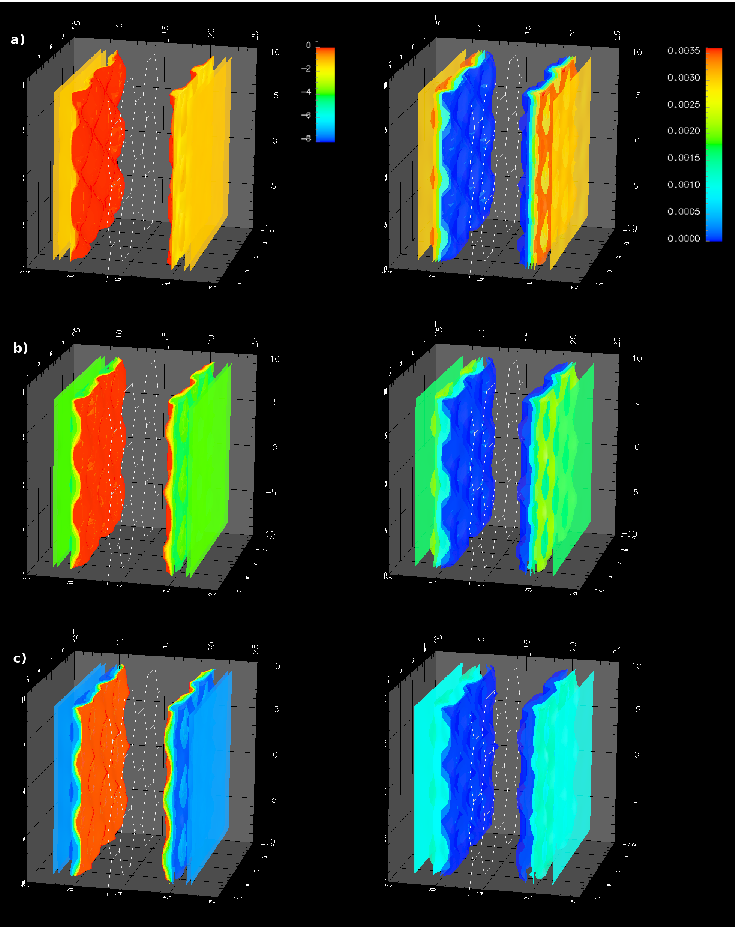}
\caption{\label{fig:VeffDens}Effective potential (left panels) and density distribution (right panels) of hydrogen adsorbed on a ZnO monolayer at a pressure $p^{(DOE)}_{~target}$=90~bar and temperature a) $T$=100~K, b) $T$=200~K, and c) $T$=300~K (energies in \kJm, density in atomic units ($a_0^{-3}$)).}
\end{center}
\end{figure}

Figure \ref{fig:VeffDens} shows the evolution of the adsorbed hydrogen density, and of the associated effective potential $v_{eff}[\rho({\vr})]$, at a higher external gas pressure ($p$=90 bar), and as a function of temperature ($T$=100~K, 200~K, 300~K). The computed density distributions confirm that at higher pressure loadings, H$_2$ molecules continue to occupy first the regions on top of zinc and oxygen atoms, and of the bonds between them.

As the temperature increases, the density profiles become gradually more homogeneous both in the perpendicular and parallel directions to the ZnO surface. The maximum local density amounts to 3.5$\cdot 10^{-3}a_0^{-3}$ at $T$=100 K. It undergoes a 37\% reduction at $T$=200 K, and an additional 22\% decrease at $T$=300 K.

The effective potentials presented in figure \ref{fig:VeffDens} exhibit an opposite behaviour:

(i) At $T$= 100 K, the attractive part of the effective potential energy surface is rather flat, i.e., spatial variability is within 1\kJm{} in the represented region.

(ii) The effective potential landscape becomes visibly more rough as the temperature gets larger. At room temperature,  $v_{eff}[\rho({\vr})]$ resembles the bare H$_2$-surface interaction potential (cf. figure \ref{fig:vext}).

The observed trend is a consequence of the intermolecular H$_2$-H$_2$ interactions being dominated by excluded volume effects. At low temperatures, it is energetically more favourable for H$_2$ molecules to occupy positions around preferential adsorption sites close to the ZnO layer, and the contribution of the short-range repulsion between piled-up molecules mask out the main features of the external potential $v_{ext}(\vr)$ to the effective potential. For increasing temperatures, there is a progressive depletion of the hydrogen uptake by the nanostructure, which renders the contribution of the H$_2$-H$_2$ repulsive forces less relevant.

\section{Conclusions}
\label{sec:summ}

Hydrogen physisorption on ZnO nanosheets was investigated within the framework of the Quantum Liquid Density Functional Theory. The density functional theory for quantum liquids at finite temperature was employed to evaluate the thermodynamic properties (e.g., isosteric heat of adsorption, volumetric and gravimetric storage capacities) and the microscopic density distribution of the adsorbed hydrogen fluid.

Analogously to other candidate materials for hydrogen storage, the enthalpy of H$_2$ adsorption to the ZnO substrate experiences a pronounced depletion for increasing temperatures. Therefore, enhanced hydrogen uptake driven by physisorption on zinc oxide monolayers is constrained to relatively low temperatures. 

The computed total volumetric and gravimetric storage capacities are very sensitive to the external gas pressure and temperature. While modest amounts of hydrogen can be stored on ZnO sheets ($n_{\rm{ads}}\le$4~kg/m$^3$, $gw\le$1\%), in the range of thermodynamic conditions specified by the US DoE targets for onboard hydrogen storage systems, the DoE goals for volumetric and storage capacities can be attained if moderately low temperatures (i.e., 77~K$\le T \le $170~K) and not too high pressures ($p\le$100~bar) are acceptable. Specifically, the target gravimetric capacity of 4.5\% is fulfilled at liquid nitrogen temperature at the ``maximum delivery pressure'' of 12~bar. 

On the one hand, the calculated storage capacities indicate a limited applicability of layered ZnO as efficient hydrogen storage media, since it can be outperformed by a variety of more complex nanomaterials. On the other hand, the computational investigation of the adsorption of molecular hydrogen on ZnO monolayers remains very appealing, since ZnO sheets can be regarded as an intermediate nanostructure between the archetypal graphene (representative of a wide class of carbonaceous nanomaterials) and MOF-5 (a prototype of the family of isoreticular metal-organic compounds). As a consequence, these nanostructured surfaces constitute a natural model system to gain further insight into the effects of the molecular geometry and of the chemical composition on the hydrogen uptake in a group of nanomaterials.

For instance, the volumetric hydrogen storage capacities on ZnO nanosheets at liquid nitrogen temperature were found to be in qualitative correspondence with previously reported theoretical and experimental results for the hydrogen uptake in MOF-5 \cite{Thomas14,kaye07,zhou07,wongfoy06,panella06,dailly06}.

Moreover, the density distribution of adsorbed molecules is clearly localized in the direction perpendicular to the ZnO surface, similarly to H$_2$ adsorbed on graphene platelets. Nevertheless, as a fingerprint of the distinct H$_2$-substrate interaction potentials for ZnO and graphene nanosheets, partially resolved adsorption peaks can be readily identified in the plane parallel to the heterogeneous substrate, opposite to the nearly uniform lateral distribution observed for hydrogen adsorption on graphitic surfaces.

In summary, owing to their high specific surface area and the enhanced H$_2$ binding strength (compared to carbonaceous materials), honeycomb ZnO monolayers exhibit competitive volumetric and gravimetric hydrogen storage capacities at cryogenic temperatures and moderate pressure, which are intermediate between those of graphene and MOF-5. While the present simulations illustrate the capability of ZnO platelets to mimic the observed trend for the hydrogen uptake by MOF-5 at liquid nitrogen temperature, the hydrogen storage properties of monolayer ZnO-based materials may be further improved by tuning their structure and chemical composition.

\section*{ACKNOWLEDGEMENTS}

The results incorporated in this publication has received funding from the European Union's Horizon 2020 research and innovation programme under the Marie Sklodowska-Curie grant agreement n$^o$898663. The work was partially supported by the Abdus Salam International Centre of Theoretical Physics within its Associate Scheme (A.M.M.), and by the Montpellier Advanced Knowledge Institute on Transitions (MAK’IT) within its Visiting Scientist programme (L.U.P.).

\end{document}